\documentclass[aps,pra,twocolumn,groupedaddress,showpacs,showkeys,superscriptaddress]{revtex4-1}

\usepackage{amsmath}
\usepackage{amssymb}
\usepackage{graphicx}
\usepackage{epstopdf}

\usepackage{subcaption}
\usepackage{mwe}

\usepackage{media9}
\usepackage{hyperref}

\begin{document}

\title{Monolithic integration of blue light sources into silicon nitride photonic chips}

\author{Ivan A. Pshenichnyuk}
\email[correspondence address: ]{i.pshenichnyuk@skoltech.ru}
\author{Muneeb Farooq}
\author{Daniil S. Zemtsov}
\author{Denis M. Zhigunov}
\author{Sergey S. Kosolobov}
\author{Vladimir P. Drachev}
\affiliation{Skolkovo Institute of Science and Technology, Moscow 121205, Russian Federation}

\date{\today}

\begin{abstract}
We investigate theoretically photonic chips with monolithically integrated blue light sources. According to our evaluations, a group-III nitride light emitting heterostructure can be efficiently combined with silicon nitride waveguiding layers. Low losses, high level of miniaturization and built-in light injection mechanism potentially make the selected platform attractive for applications and draw up an addition or even alternative to silicon photonics.
We use a combination of drift-diffusion and Maxwell equations to build a model of the proposed multilayer structure. The model allows to choose the best parameters for both light emitting sandwich and waveguiding layers as well as to pick an optimal coupling regime. High transmittance coefficients are obtained. Various optimal geometries are analised with respect to captured power, desired polarization and excited modes. The obtained numbers are promising and allow in the future to pave the way towards hetero-integrated blue photonic circuits.
\end{abstract}

\maketitle

\section{Introduction} \label{sec_intro}

Silicon based integrated photonics represents a perspective direction of the future technological development \cite{siew-2021}. Recent advances in the area of active photonic devices  \cite{zemtsov-2023, pshenichnyuk-2019, pshenichnyuk-2018c} and the implementation of advanced plasmonic switching mechanisms \cite{pshenichnyuk-2021, pshenichnyuk-2024} allow to expect an appearance of compact and efficient integrated photonic chips in the future. But there is still a long way in making this technology universal and mature. One of the problems waiting for an appropriate solution is the integration of light sources and chips. The most used way to power photonic integrated circuits (PIC) is to couple a radiation via optical fibers. Different coupling schemes are used including grating couplers and edge couplers \cite{marchetti-2019,zemtsov-2022}. Any of those schemes are bulky compared to the size of the chip itself. It is desirable to have a source of light monolithically integrated to obtain self-reliant chips. A laser or light emitting diode (LED) heterostructure should allow coupling emitted light into an underneath integrated waveguide with high efficiency. Even though some attempts are made \cite{komljenovic-2018} the general problem remains unsolved. 

One should first choose an appropriate light emitting heterostructure and a suitable low loss guiding material. Note, that light emitting sandwiches usually can guide it themselves, but they are lossy and not suitable for large scale chip building. Thus, it is necessary to separate emitting and guiding layers, providing an efficient conversion mechanism.
The effective indices of both components should be comparable to ensure efficient near-field based conversion at the length scale of a chip. Usually, it means that the refractive indices of waveguiding materials should have similar values.
Both chip platform materials and heterostructure materials should be physically linkable, directly, or using some kind of intermediate material layers allowing to overcome a probable crystal lattices mismatch. The mismatch spoils the purity of light emitting crystals and makes them impractical. The problem is in some sense similar to one of the problems solved by Shuji Nakamura trying to create blue LED \cite{akasaki-2007}. 
Important steps in the technology of monolithic integration were done previously \cite{wu-2003} \cite{parlinskawojtan-2015}.  

It is worth mentioning that the wavelength chosen for integration plays a significant role in the design. The usual choice for silicon photonics is $1.5$ microns, dictated by the transparency window of silicon. But it is not necessary the best choice for the integration. Moreover, considering the diffraction limit $\lambda/2$ it is desirable to use smaller wavelengths, since it allows to make photonic devices smaller. The greater size of photonic elements compared to electronic elements is a big issue in integrated photonics. From this point of view, it is preferable to go to smaller wavelengths and switch from Si to something working in the visible or even in the ultraviolet range \cite{west-2019}. Of course, there exist a limiting factor, since at UV frequencies most of the materials become absorptive.

In this paper we investigate theoretically a coupling scheme where group-III nitride light emitting platform is used to power Si$_3$N$_4$ based photonic chip.
Obvious advantages of Si$_3$N$_4$ are the following. It is a low loss material with a large transparency window, that includes visible spectral range \cite{sacher-2019} \cite{baets-2016}. Silicon nitride based photonics is CMOS technology compatible and provides a certain fabrication flexibility. It has lower refractive index as compared with Si which is an advantage in our case of the inter-layer matching.
On the other hand, blue light emitting GaN heterostructures and corresponding fabrication technologies are well developed \cite{li-2020}.  The minimalistic sandwich includes doped AlGaN and InGaN layers. The obtained LEDs are bright and efficient. The wavelength provided by blue LED around $450$ nm is convenient for integration according to our evaluations. Moreover, compared to Si infrared photonics the chosen wavelength allows to make components smaller up to the factor of $3$.
Blue light emitted by such LEDs well fits the Si$_3$N$_4$ transparency window. An efficient light extraction from the GaN sandwich to redirect it into low loss silicon nitride waveguide requires matching their refractive indices. This is a general criterion if we want to perform an evanescent wave-based transformation implied here. This condition is fulfilled for the selected groups of materials. Note, that Si is less efficient from this point of view.

Some technological efforts will be required providing the interface between In/Al/Ga nitride and silicon nitride layers to obtain hetero-integration of sufficient quality \cite{kamei-2020}. Our theoretical model allows adding a transient layer for smoothing probable mismatch in the crystal lattices if necessary. The link between two materials is experimentally realized and described in the literature \cite{wu-2003} \cite{parlinskawojtan-2015}
According to our evaluations, an affinity of the both families of materials is good enough and potentially allows to expect fully integrated self-powered blue PICs in the future.

Other advantages of the shorter wavelength are potential applications of the visible light circuits. That includes
biochemical sensors \cite{polstein-2012} \cite{li-2017},
opto-genetics \cite{hoffman-2016} \cite{shim-2016},
optical interconnects \cite{brubaker-2013},
visible range communications \cite{xie-2019},
bio imaging and quantum computations \cite{jagsch-2018},
and more \cite{song-2019} \cite{romerogarcia-2013}.
Despite the perspectives, not many attempts are made.
The monolithic evanescent integration is demonstrated experimentally for group-III nitride (GaN and AlN) on silicon platform \cite{tabatabavakili-2018,tabatabavakili-2019}. A disk-shaped blue LED is coupled to a waveguide, made from the same heterostructure materials (not silicon nitride), but without an active layer of quantum wells.
Different coupling scenarios between a silicon nitride waveguide and a blue laser are demonstrated in \citet{arefin-2020}. But the emphasis is made on the far field coupling. 

\begin{figure}
\centerline{\includegraphics[width=0.5\textwidth]{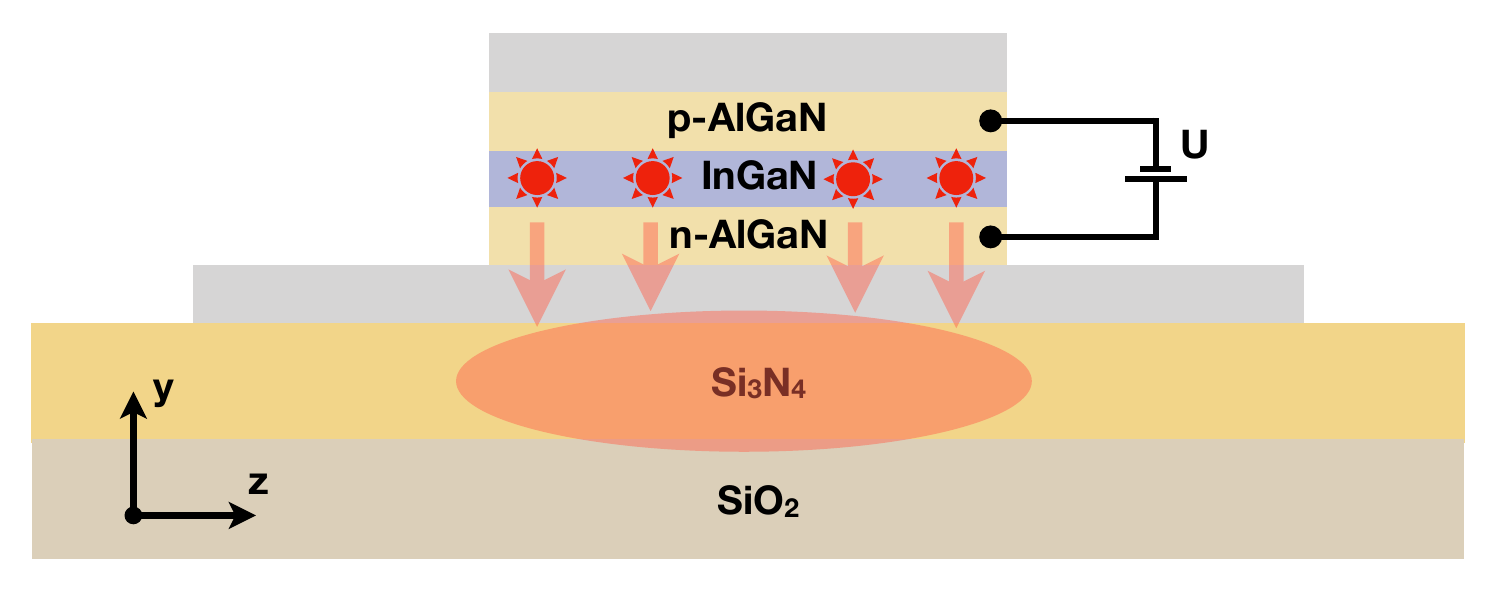}}
\caption {The general scheme of the proposed device. Light is generated in LED sandwich and coupled into lower lying low-loss waveguide with high conversion efficiency.
\label{fig1_cscheme}}
\end{figure}

In this work we consider the model scheme depicted in Fig.~\ref{fig1_cscheme}. The proposed multilayer structure is assembled from light emitting InGaN/AlGaN layers deposited on top of Si$_3$N$_4$/SiO$_2$ sandwich. Light is generated in the InGaN well caused by the applied voltage $U$. Then it is captured by the structure and transformed into specific waveguide mode inside the lower lying low-loss Si$_3$N$_4$ layer. The geometry and composition of materials should ensure the maximal possible light generation and conversion efficiency. To build an appropriate theoretical model we use a combination of the drift-diffusion equations with optical transitions and Maxwell equations. The detailed description of the approach is presented in Sec.~\ref{sec_theory}. Then, in Sec.~\ref{sec_results} we discuss the computation results including the optimization procedure and obtained parameters in different coupling scenarios. The main results are summarized in the conclusion.

\section{Theory} \label{sec_theory}

A system of the drift-diffusion equations applicable for the numerical description of light-emitting heterostructures can be derived from the more general Boltzmann's formalism \cite{rosencher,brennan}. The system consists of the continuity equations for electrons and holes 
\begin{equation}
  \frac{\partial n}{\partial t} - \frac{1}{e} \nabla \cdot \mathbf{J}_n = G-R,
  \label{dd_gen_1}
\end{equation}
\begin{equation}
  \frac{\partial p}{\partial t} + \frac{1}{e} \nabla \cdot \mathbf{J}_p = G-R,
  \label{dd_gen_2}  
\end{equation}
combined with corresponding Ohm's and Fick's laws in the form
\begin{equation}
  \mathbf{J}_n = e\mu_n n\mathbf{E} + eD_n \nabla n,
   \label{dd_gen_3} 
\end{equation}
\begin{equation}
  \mathbf{J}_p = e\mu_p p\mathbf{E} - eD_p \nabla p.
    \label{dd_gen_4}
\end{equation}
Here $n=n(\mathbf{r},t)$ and $p=p(\mathbf{r},t)$ are electrons and holes densities, $\mathbf{J}_n$ and $\mathbf{J}_p$ - corresponding current densities. Mobilities and diffusion coefficients for electrons and holes are denoted as $\mu_n$, $\mu_p$, $D_n$ and $D_p$ respectively. The system is coupled with Maxwell's equations through the electric field $\mathbf{E}$. Generation and recombination rates are given by $G$ and $R$.

Assuming that the electrons and holes dynamics is much slower that the field we may consider an electrostatic approximation and reduce the system of Maxwell's equations to the following Poisson's equation \cite{rosencher}
\begin{equation}
  \nabla\cdot (\varepsilon\nabla\varphi)=e(n-p+N_A-N_D)
   \label{dd_1}  
\end{equation}
for the potential $\varphi=\varphi(\mathbf{r},t)$.
Here we introduce the medium permittivity $\varepsilon$ and densities of immovable donors and acceptors $N_D$ and $N_A$.

The system can be simplified further using the following statements. When only the steady state is analyzed the time derivatives can be eliminated. We also consider $G=0$, as there are no internal sources of electrons and holes. Then from Eqs.~\ref{dd_gen_1}--\ref{dd_gen_4} we obtain
\begin{equation}
  \nabla\cdot[D_n\nabla n - \mu_n n\nabla(\varphi+\chi)]=R,
   \label{dd_2}  
\end{equation}
\begin{equation}
  \nabla\cdot[D_p\nabla p + \mu_p p\nabla(\varphi+\chi+E_g)]=R.
   \label{dd_3} 
\end{equation}
Here we considerr that the potentials in the valence and conduction bands are shifted by the values of the bandgap energy $E_g$ and electron affinity $\chi$. These quantities undergo steplike variation at the boundaries between materials. The system of equations \ref{dd_1}--\ref{dd_3} is self contained and prepared for the numerical treatment. As a result we obtain the steady state electron/hole distribution functions and potential: $n(\mathbf{r})$, $p(\mathbf{r})$, $\varphi(\mathbf{r})$.

The recombination rate $R$ in general includes various terms that correspond to different recombination mechanisms. In this work we take into account Auger recombination in the form
\begin{equation}
  R_a = (C_nn+C_pp)(np-n_e^2),
\end{equation}
where $C_n$ and $C_p$ are Auger coefficients and $n_e$ effective concentration. At the same time, we neglect other non-radiative transitions, proportional to the first power of concentration, assuming them to be much smaller in the considered regimes.

The spontaneous emission rate is given by the equation
\begin{equation}
  R_{sp} = \int \tau^{-1} f_c(1-f_v) g_{red} dE.
  \label{r_ph_sp}
\end{equation}
Here $\tau$ is the radiation lifetime inversely proportional to the recombination probability, $g_{red}$ -- reduced density of states, $f_c$ and $f_v$ - populations in the conduction and valence bands respectively. And the integral is taken through all possible transition energies. Spontaneous emission can occur at any frequency allowed by the band structure and the spectrum is in general broader compared with the stimulated emission. The phase of spontaneous emission is random. $R_{sp}$ implicitly depends on coordinates through the positions of quasi Fermi levels $E_{F_c}$, $E_{F_v}$ and defines the spatial distribution of luminescence \cite{rosencher}. 

In the calculations we keep LED in the electroluminescence regime, below the lasing threshold. It is guaranteed by the Bernard-Durraforg criterion
\begin{equation}
  E_{F_c} - E_{F_v} < E_g,
   \label{bd_crit} 
\end{equation}
which is fulfilled in all the regimes considered below. This restriction allows to exclude the stimulated emission $R_{st}$, which is not that important for the optimization tasks we solve below. So, the emitted light is incoherent. From the theory point of view, the presence of stimulated emission does not allow to decouple Maxwell and Drift Diffusion equations, since $R_{st}$ depends on the intensity of light. Thus, the exclusion of stimulated emission provides a significant computational advantage.

We use the frequency domain wave equation \cite{pshenichnyuk-2019}
\begin{equation}
  \nabla\times\nabla\times\mathbf{E}(\mathbf{r}) - k_0^2n^2 \mathbf{E}(\mathbf{r}) = 0
  \label{wave_equation}
\end{equation}
to calculate the propagation of light emitted via the recombination process. Here $n$ is a complex refractive index of a medium. The source is modelled using oscillating electric dipoles. Since the spontaneous emission in semiconductors is known to be isotropic \cite{rosencher} we consider all possible directions and polarizations as equiprobable. Three different dipoles oriented along main orthogonal coordinate axis form a suitable basis for the representation of an arbitrary source. In general, it can also be used to build semi analytical formalisms like the Green's functions approach \cite{novotny}.

The combination of the drift-diffusion and Maxwell equations allows to describe both electronic and optical characteristics of an integrated light source. That includes the band structure modification under the influence of voltage, current-voltage characteristics, threshold current, a spatial distribution of emission intensity, LED optical spectrum and many other characteristics. Going back to this work it allows to pick the optimal integration parameters for both light-emitting heterostructure and underlying waveguiding subsystem. The optimization includes the variation of geometrical parameters and doping profiles that influence both light-emitting characteristics and transformation efficiencies.

The suggested model is parametrized using the data published in literature \cite{zhang-2009} as well as Comsol material library. The composition of light-emitting sandwich is taken as In$_{0.126}$Ga$_{0.874}N$ for the $5$ nm thick central layer and Al$_{0.15}$Ga$_{0.85}N$ for claddings. The width of claddings is varied in computations. The concentration of donors and acceptors in n-doped and p-doped areas is $N_D=N_A=10^{18}$ cm$^{-3}$. Gap energies $E_g$ for InGaN and AlGaN are $2.76$ V and $3.7$ V, while electron affinities $\chi$ are $4.6$ V and $4.1$ V respectively. Mobilities of electrons and holes are $\mu_n=1000$ cm$^2$V$^{-1}$s$^{-1}$, $\mu_p=350$ cm$^2$V$^{-1}$s$^{-1}$. The refractive index $n$ is taken as $2.6$, $2.4$, $1.5$ and $2.1$ for InGaN, AlGaN, SiO$_2$ and Si$_3$N$_4$ respectively. Auger recombination coefficients are $C_n=C_p=1.7\times10^{-30}$ cm$^6$/s and the recombination time is $\tau=2$ ns. Numerical computations are made using Comsol Multiphysics 5.3a software with additional modules Semiconductor and Wave Optics. Lenovo ThinkStation P620 hardware is used to run the simulations.

\section{Results and Discussion} \label{sec_results}

\begin{figure*}
\centerline{\includegraphics[width=1.0\textwidth]{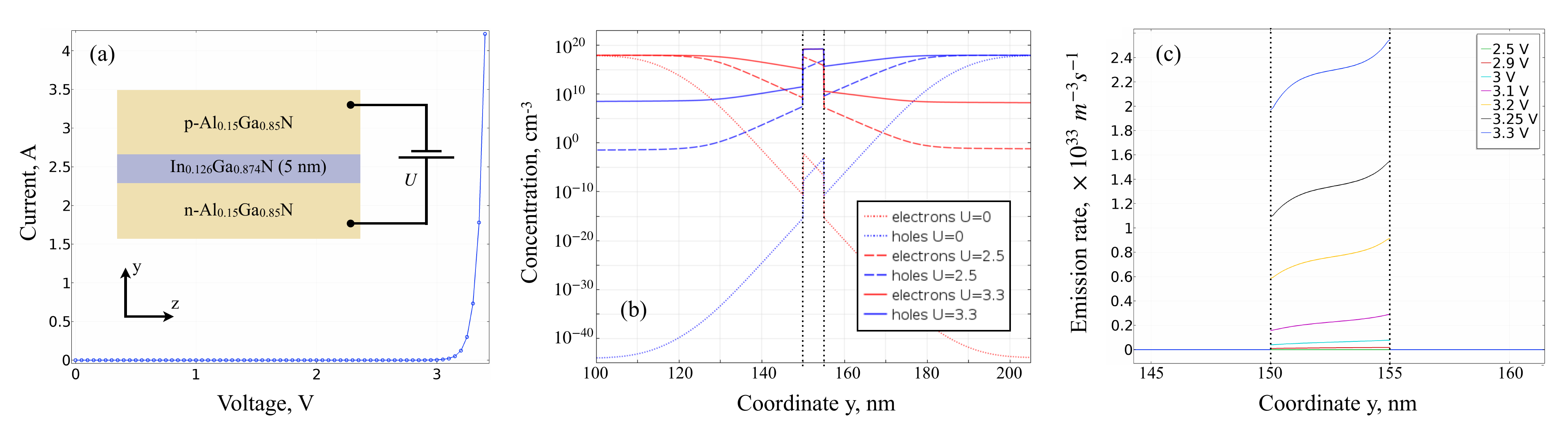}}
\caption {The results of the light-emitting sandwich electronic properties modeling: (a) the current-voltage characteristic, (b) electrons and holes density distribution, (c) the spatial distribution of electroluminescence.
\label{fig2_ddresult}}
\end{figure*}

First, we present the results of electronic computations for the light-emitting sandwich, as depicted in Fig.~\ref{fig2_ddresult}. The conductance and light emitting profiles are investigated along $y$-axis. The width of the central InGaN layer is fixed at $5$ nm. This layer forms a well, designed to maximize light emission, and increase in its width causes drops in LED brightness. But the width of AlGaN claddings is the subject of variation as will be discussed below. Theory presented in the previous section allows to compute the current-voltage characteristics and determine the lasing threshold according to Eq.~\ref{bd_crit} that appears at approximately $3.3$ volts here. Below this voltage the LED works in electroluminescence regime, where spontaneous emission dominates.

The distribution of densities of electrons and holes at different values of voltages are shown using the logarithmic scale in Fig.~\ref{fig2_ddresult}b. At zero voltage the concentration of electrons and holes in the well is quite small. And the concentration of corresponding carrier types in n-doped and p-doped areas of the junction is defined by the selected values of doping $N_D$ and $N_A$. The concentration of electrons and holes in the well grows with voltage and gets close to $10^{20}$ cm$^{-3}$ for voltages around $3.3$ V, creating favourable conditions for the recombination. The corresponding emission profiles are shown in Fig.~\ref{fig2_ddresult}c for different voltages. Note, that electrons and holes have different effective masses and mobilities. That results in the absence of full symmetry between their density profiles, especially inside the well (Fig.~\ref{fig2_ddresult}b).

The light emitting sandwich with refractive indices $2.6$ and $2.4$ for the core and claddings respectively, being surrounded by materials with lower indices, like for example, SiO$_2$ with $n=1.5$, forms a waveguide. The first task of the integration is to capture as much of the emitted energy as possible inside this waveguide. Note, that the task is opposite to the one solved for ordinary LED, where it is desirable to emit as much power as possible into the far field. We also note that the waveguide formed by doped materials is not good for optical guiding, since it introduces significant losses, caused by the increased amount of charge carriers. So, the second necessary step is to convert the trapped power into the low loss waveguiding layer mode. 

The problem of LED beam engineering and light extraction is often solved using ray tracing techniques \cite{piprek,lalaukeraly-2017}. Here, taking into account the compact geometries and intense usage of near fields we implement the wave optics-based approach (Eq.~\ref{wave_equation}). The emission of light takes place in InGaN layer with the emission profile depicted in Fig.~\ref{fig2_ddresult}c. Since the emission layer is quite thin compared to the emitted wavelength and the thickness of other layers, the precise position of each source caused by some recombination event does not influence much the optimization parameters. It allows us to apply averaging and assume that the dipole is always located in the middle of InGaN layer. 
The wavelength of emitted light that corresponds to the selected value of band gap $E_g$ is $450$ nm. At the same time the spectrum of possible transitions in the integral of Eq.~\ref{r_ph_sp} depends on the voltage. Close to $3.3$ V at room temperature the maximum of emission shifts slightly to $430$ nm. This value is used for the optical modeling. 
 
\begin{figure}
\centerline{\includegraphics[width=0.5\textwidth]{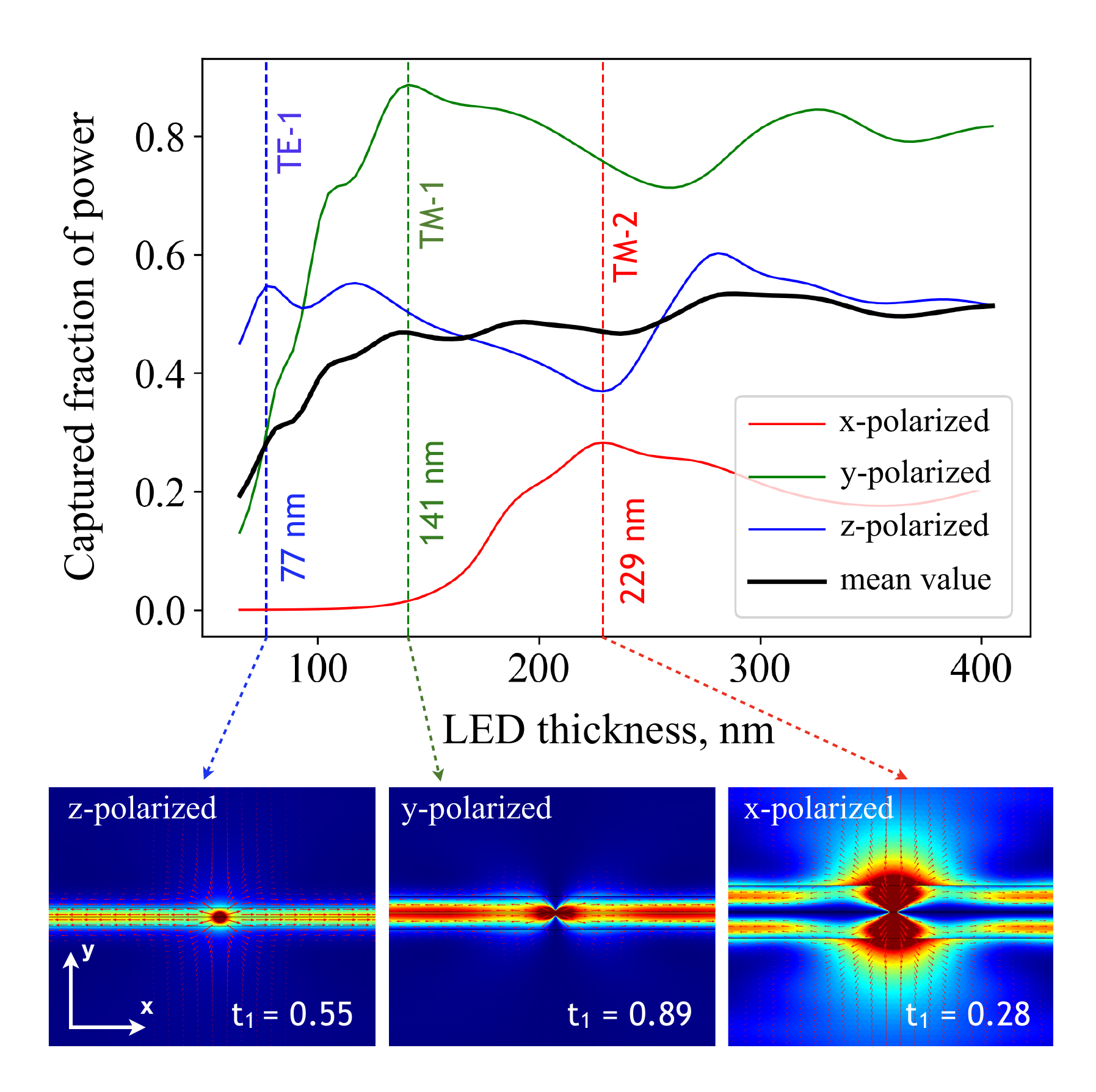}}
\caption {Optimization of LED thickness for three different polarizations of the emitted light, shown by red, green and blue curves. Three optimal coupling scenarios corresponding to the excitations of different modes are marked by vertical dashed lines. Corresponding modes are visualized using the power flux spatial distribution. An average value is shown using the black curve. 
\label{fig3_ledopt}}
\end{figure}

The dependence of captured fraction of power on LED thickness (both InGaN and AlGaN layers) is shown in Fig.~\ref{fig3_ledopt}. The LED is assumed to be surrounded by SiO$_2$ from both sides, but other materials with similar refractive indices are possible. Three different polarizations of light are considered to cover all possible emission scenarios. The structure is assumed to be sufficiently wide in $z$-direction ($\sim 10$ $\mu$m and more). It guarantees that slab modes are formed inside the waveguide and 2D modeling is sufficient to describe them.

The captured fraction of power depends on the LED thickness in a nontrivial way, as depicted in Fig.~\ref{fig3_ledopt}. Multiple maxima correspond to the excitation of different waveguide modes. And the process, obviously, strongly depends on the polarization. The results for $x$-, $y$- and $z$-polarized sources are shown by red, green and blue curves respectively. First strong maxima appear at thicknesses $229$ nm, $141$ nm and $77$ nm. In these three coupling scenarios different modes are predominantly excited, namely TM-2, TM-1 and TE-1. The relative distribution of power flux absolute value and its direction in these 3 cases is demonstrated in Fig.~\ref{fig3_ledopt} (lower panel). The corresponding maximal transmission coefficients $t_1$ are $0.28$, $0.89$ and $0.55$.
The thickness of choice depends on the subsequent application of the integrated light source. If the goal is to excite some specific mode with maximal possible efficiency, then one of the scenarios should be picked, taking into the account $t_1$ coefficient. If the goal is to pump maximal power into the waveguide, and the mode composition is irrelevant, an averaged curve should be taken. Assuming that all three polarizations of the source are equiprobable, the mean value is demonstrated by black curve in Fig.~\ref{fig3_ledopt}. In this case the optimal thickness is close to $300$ nm.

When the emitted light is trapped inside the emission layer, the second step is to transform it into the low loss waveguide for subsequent usage inside the integrated chip. The most elegant mechanism is to perform transformation via near field coupling, as implemented in directional couplers. It is known that the condition for highly efficient transformation is the similarity between effective mode indices in both waveguides \cite{yariv}. Effective indices depend on the materials and geometry of a waveguide, as well as polarization and mode number. Since both waveguides are made from different materials with different refractive indices, the matching can be realized using a variation of the geometry.

\begin{figure}
\centerline{\includegraphics[width=0.5\textwidth]{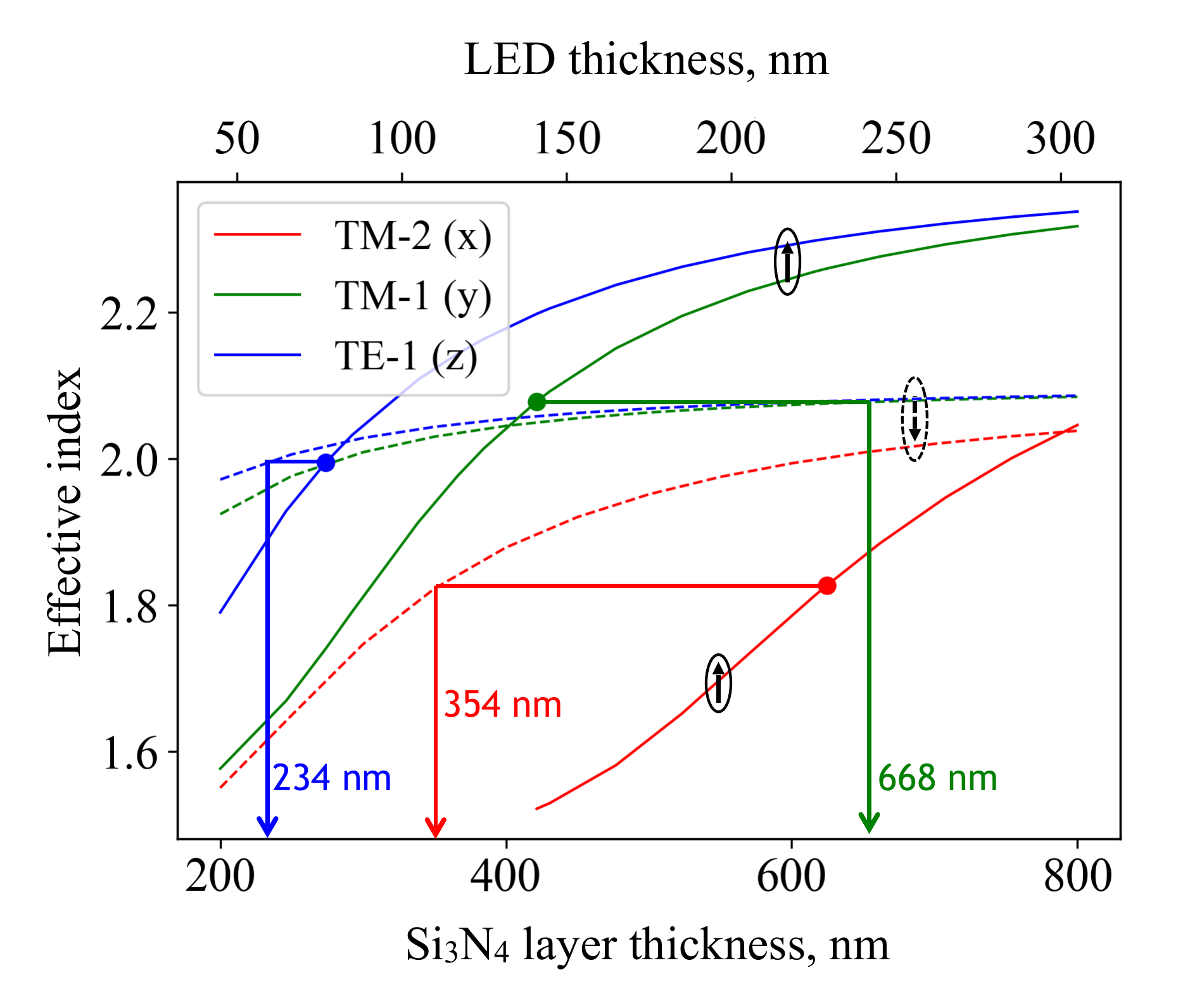}}
\caption {Visualization of the effective indices matching procedure between LED sandwich mode and Si$_3$N$_4$ waveguide mode. Three different modes that correspond to different coupling scenarios are considered.
\label{fig4_neffsync}}
\end{figure}

The matching procedure is visualized in Fig.~\ref{fig4_neffsync} independently for three different coupling scenarios described above. Different polarizations excite different modes, with their own effective indices, that also depend on LED thickness. These dependences are given by red, green and blue solid curves. The optimal thickness determined above is shown using bold spots of analogous colors.
The modes of Si$_3$N$_4$ waveguide with corresponding symmetry and polarization also depend on the waveguide thickness. They are shown using dashed curves of similar colors. The optimal width corresponds to the value of effective index equal to the one denoted by the bold spot of the corresponding color. So, we simply draw the horizontal lines that connect spots with dashed curves of the same color (Fig.~\ref{fig4_neffsync}). Solutions exist in all three cases. The obtained values of Si$_3$N$_4$ thickness for three coupling scenarios are $354$ nm, $668$ nm and $234$ nm.

\begin{figure}
\centerline{\includegraphics[width=0.5\textwidth]{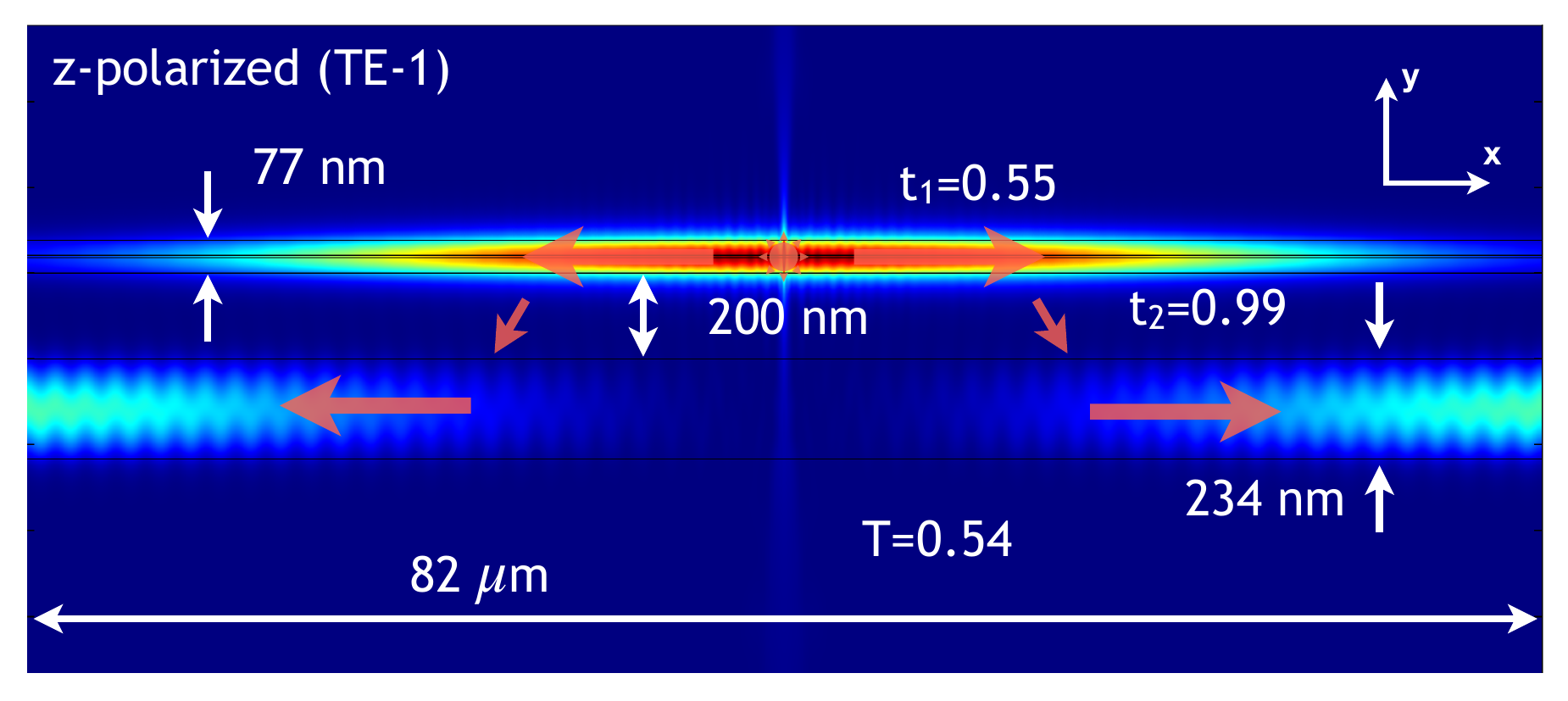}}
\caption {The complete optimized optical model that shows how the light is emitted by LED, trapped (with some losses) inside the light-emitting sandwich and then transformed into Si$_3$N$_4$ low loss mode. The process is demonstrated for the TE-1 mode.
\label{fig5_fullmodel}}
\end{figure}

To determine the coupling efficiency the whole system should be explicitly modeled. It allows us to determine the second stage conversion efficiency coefficient $t_2$, as well as the whole device efficiency coefficient $T=t_1t_2$. A well-designed directional coupler optimized according to the algorithm above gives near unity coefficient $t_2$. To be precise, for three scenarios described above we obtain values $0.96$, $0.98$ and $0.99$. So, most of the losses appear at the first level of conversion. The corresponding coefficients $T$ are $0.27$, $0.87$ and $0.54$.
The distribution of power flux density in one of the scenarios, namely TE-1 mode, is shown in Fig.~\ref{fig5_fullmodel}. The light emitted in InGaN layer is trapped in the first waveguide formed by AlGaN/InGaN sandwich and then transformed into the underlying low loss Si$_3$N$_4$ waveguide.

It is important to note that along with the coupling efficiency $t_2$ the directional coupler has other important characteristics such as the conversion length and the gap size. Smaller gaps correspond to larger coupling between waveguides and shorter coupling distances. It is preferable to keep coupling distances short to decrease the longitudinal size of the device. At the same time large coupling makes the conversion process harsher. With more losses and excitation of undesired modes. Moreover, coupling modes theory is formulated for weak coupling regime and can work poor for small gaps. So, there must be a good compromise between these two parameters, depending on subsequent applications of the chip.

In Fig.~\ref{fig5_fullmodel} we present a device with $200$ nm gap that corresponds to $41$ $\mu$m coupling length. It should be doubled if we capture the light that is emitted in both directions. We also note that the excited mode is not completely pure. Waveguide thickness $234$ nm allows the existence of multiple TE modes, and some energy goes there during the conversion. The purity of the mode can be further increased for larger gaps and more gentle conversion procedure. Complete purification of the output mode can be achieved using a single mode waveguide. That would require the usage of a different wavelength and alternative materials. For the present study, where the emitted light is considered as incoherent, it is not that important. This problem can be considered in future works along with more general theory that goes beyond electroluminescence regime.

\section{Conclusion} \label{sec_conclusion}

We have investigated theoretically the possibility to build a photonic chip with a monolithically integrated light source using a silicon nitride platform combined with group-III nitride light emitting materials. It is demonstrated that such chips can operate at the visible light around $450$ nm, creating an alternative to more traditional $1.5$ $\mu$m silicon photonics and promising additional miniaturization of devices. The presented computational algorithm includes the combination of drift-diffusion equations and Maxwell equations. It can be used for the optimization of self-sufficient light emitting chips. It is demonstrated that the mode matching task is solvable for the selected groups of materials, which is not a priori obvious. Different coupling scenarios characterized by high light conversion efficiencies are demonstrated and discussed. The existing compromises between the parameters that should be resolved according to the desired applications of devices are discussed.

\begin{acknowledgments}
We thank Prof. Nikolay Gippius and Prof. Yury Gladush from Skoltech for fruitful discussions.
\end{acknowledgments}

\bibliography{paper_blueled}

\end{document}